# Identification of Mechano-stimuli (Linear to Harmonic) Thermal Response in Mesoscopic Liquids

*Eni Kume[1] and Laurence Noirez[1]\**

[1]Laboratoire Léon Brillouin (CEA-CNRS), Univ. Paris-Saclay, 91191 Gif-sur-Yvette Cedex, France

KEYWORDS: mesoscopic scale, liquid, low frequency excitation, wetting, thermoelasticity.

ABSTRACT: In the conventional picture, the temperature of a liquid bath in the quiescent state is uniform down to thermal fluctuation length scales. Here we examine the impact of a low frequency shear mechanical field (Hz) on the thermal equilibrium of liquids (Polypropylene glycol and pentadecane away from any phase transition) confined between high energy surfaces. We show the emergence of both cooling and heating shear waves of several tens microns widths varying synchronously with the applied shear strain wave. The thermal wave is stable at low strain amplitude and low frequency while thermal harmonics develop by slightly increasing the frequency or the strain amplitude. The liquid layer behaves as a thermoelastic medium. This view is in agreement with recent theoretical models predicting that liquids support shear elastic waves up to finite propagation length scale of the order the thermal wave.

## 1. Introduction:

The mechanical properties of fluids are extensively studied in terms of stress response but not in terms of thermal response. Mechanically induced thermal effects are typically solid-like properties, known as thermoelasticity. For small changes of temperatures, the solid thermoelastic effects are generally weak, reversible and giving rise to an adiabatic transformation where the compressibility/dilatation ability of the material is varied under mechanical stress: $\Delta T \propto -\frac{\alpha_L}{C_\sigma}.\sigma$ where $\sigma$ is the stress component, $\alpha_L$ the linear thermal expansion and $C_\sigma$ the specific heat at constant stress. This thermal variation can be positive or negative depending on the nature of the transformation leading to stretched or compressed regions respectively [1]. Without heat source, no temperature variation is predicted for fluids upon shear mechanical action as the energy is expected to be entirely dissipated on large volumes [2]. The situation might be different by exploring low scale liquid behaviour. Indeed while the liquid definition implies an absence of static shear elasticity, shear elasticity has been recently predicted and experimentally measured in mesoscale liquids [3-7]. Volino proposed a non-extensive model to describe dimension-dependent liquid viscoelasticity [8]. Revisiting the Frenkel model, Trachenko et al. introduce the concept of local dynamic compressive stress and revisits the Maxwell approach showing that a solid-like approach can be treated on equal footing introducing the notion of finite shear wave propagation length (gapped momentum states) [9-13]. These models converge in interpreting the macroscopic liquid behavior as an asymptotic branch where, at a smaller scale, liquid molecules behave elastically with suppressed relaxations and where the dynamic role of intermolecular interactions is central [12-13]. Since evidence of small-scale elasticity seems established, it is natural to test whether another solid property, thermoelasticity, is also observable. Thermal variation involves an energy change between molecules through short time dependent interactions of the vibrations of the molecules. The far-infrared portion of the wide electromagnetic spectrum covers wavelengths from approximately 3 to 14 µm. This wavelength range concerns typically the intermolecular vibrations of the liquid dynamics; i.e. it characterizes the way molecules vibrate in the "quasi-cage structure" of the neighbored molecules. Except for solid stress analysis [14], this fundamental property that is the thermal radiation naturally emitted by the condensed matter above zero degree Kelvin is not exploited to access the liquid thermodynamic state. We present a high accuracy real-time micro-thermal analysis of a liquid layer of a few hundred of microns set between highly wetting surfaces under oscillatory shear deformation conditions. The studied liquid is the glass former polypropylene glycol (PPG) at room temperature; i.e. far from any phase transition. This liquid is nearly transparent to infra-red wavelength so that the bulk behavior can be probed. We show that the liquid responds thermally to the low frequency mechanical excitation challenging the assumption of fast thermal relaxation in liquids. We characterize the thermal signal and identify thermal harmonics when the excitation becomes larger in magnitude and frequency. Preliminary tests were also carried out on single molecule non polar solvent, the pentadecane (alkane), allowing us to pretend that the following results might be representative of a generic liquid property.

## 2. Method:

For the experiments, two ordinary liquids, the polypropylene glycol (PPG, Sigma-Aldrich $M_n$ = 4000) which is a H-bond glass former with an extremely low evaporation rate, and the pentadecane (Sigma-Aldrich) which a molecular van der Waals liquid were used. The measurements took place at room temperature ~20°C, which is far from the phase transition of PPG ($T_g$ = 180 K) and pentadecane (melting temperature: 282.3 K [15]).

The liquids were confined in between two parallel 50mm diameter α-Alumina plates that guarantee a total liquid/solid wetting [4-6]. The plates were cleaned by heating for an optimal liquid/substrate wetting ensuring ideal conditions for the transmission of a dynamic shear strain generated from the sinusoidal movement of the bottom plate $\gamma = \gamma_0 . sin(\omega.t)$ where $\gamma_0$ is the strain amplitude and $\omega$ the frequency. Imposing a sine wave deformation allows by changing the frequencies and amplitudes to examine the stability of the signal response. A strain-controlled rheometer (TA-Instruments ARES) was used to impose the frequency, the strain amplitude and the gap thickness. A Keithley multimeter is coupled as analog output to the rheometer and to a PC to allow the synchronization with an infrared microbolometer camera. The camera matrix of 388 * 282 pixels of 20 µm uses wavelengths from 8 to 14 µm at a frequency reading of 27 Hz, making possible the observation of the thermal response during a full mechanical oscillation since the applied frequency varies from 0.5 to 5 rad/s (0.08 to 0.8 Hz). The camera lens was placed at 50mm away from the plate-liquid-plate circumference focusing at a depth of ~0.85mm ± 0.1mm (depth of field) inside the liquid, probing the liquid bulk. The infrared data were corrected by subtracting the average thermal frame recorded at the equilibrium from every frame. The kinetic representation of the thermal mapping was created by the compacted succession of the frames.

## 3. Results

### 3.1 General features:

Fig.1a and b show the temperature variation in the gap recorded along two oscillatory periods of the liquid PPG at $\omega$ = 0.5 rad/s and 1 rad/s respectively (at 240µm gap thickness and $\gamma_0$ =4000 %). This kinetic representation of the temperature in the gap reveals a succession of instant thermal states at different stages of the oscillatory deformation. The different colors in Fig.1a and Fig.1b indicate that the liquid undergoes a thermal change with both local negative and positive temperature variations that reach sizeable amplitudes of about ± 0.2° C. These thermal waves are reproducible, reversible and stable in time as long as the mechanical strain wave is applied. The coexisting hot and cold waves are roughly distributed in three bands (of ~ 50µm thickness each) that we name "bottom, middle and upper bands". The bands exhibit in phase and opposite thermal waves which compensate for each other; Fig.1c and 1d illustrates the temperature variation in each chosen bands. Cold and warm zones alternate synchronously with the applied excitation; i.e. while the central band is cooling down, the neighbored ones are heating up and reciprocally.

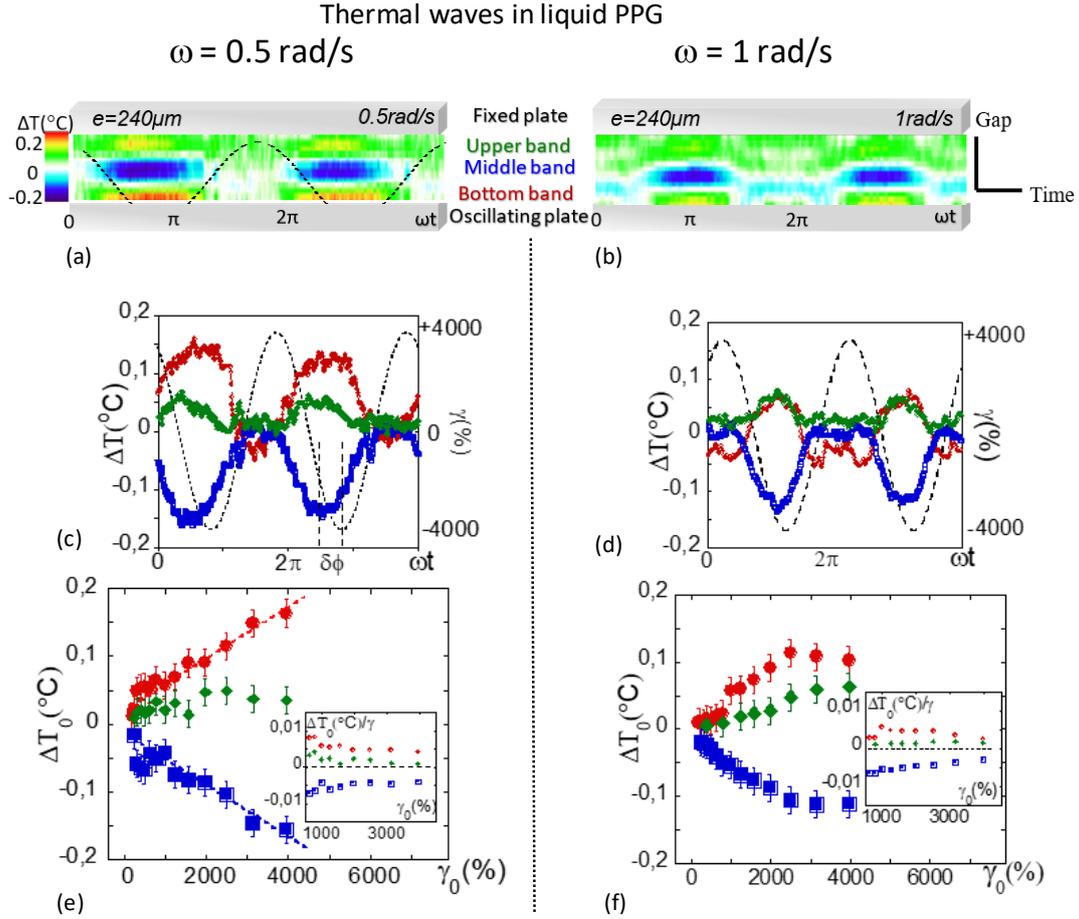

**Figure 1.** By applying a low frequency mechanical stimulus (~ Hz), the liquid emits a modulated thermal signal synchronous with the stimulus (here PPG-4000 confined in a 240μm gap (gap view), alumina substrate, room temperature measurements). **a)** Real-time mapping of the temperature γ = 4000% (gathering about ~800 frames): at ω = 0.5 rad/s **b)** same at ω = 1 rad/s. The dotted line is an eye guide for the applied strain. **c)** Thermal waves recorded for the bottom (●), middle (■) and upper (♦) bands respectively at γ = 4000% and ω = 0.5 rad/s. The dotted line illustrates the applied shear strain. δφ is the phase shift between the thermal and strain wave (δφ ≈ 45° at γ = 4000%). **d)** Same for ω = 1 rad/s. **e)** Strain dependence of the maximum of the temperature variation ΔT$_0$(°C) versus shear strain and inserts representing the "gain" ΔT/γ versus shear strain for the bottom "hot" band: (●), the middle "cold" band: (■) and the upper band: (♦) respectively at ω = 0.5 rad/s, **(f)** same at ω = 1 rad/s.

## 1.2 Strain dependence of the thermal signal:

The strain dependence of the cold wave is shown in Fig.1e and Fig.1f at 0.5 rad/s and 1 rad/s respectively. At low strain values (< 500%), the signal of thermal response is weak and noisy being close to experimental error bar (~ 0.02°C) and making difficult to analysis of the thermal response at low strain amplitude.

At 0.5 rad/s, the strain dependence of the main waves (hot and cold) is relatively simple exhibiting a nearly linear dependence over a large strain range (Fig.1e). The thermal modulation can be modelled by a sin wave: $\Delta T = \Delta T_0 \cdot \sin(\omega.t + \delta\varphi)$ where $\delta\varphi$ is the phase shift that is nearly in phase at low strain (γ < 1000% not shown), and above, reaches approximately π/4. The sin wave is kept up to the highest reachable strain values (Fig.1c is obtained at 4000%), indicating the additional energy associated to the increase of the shear strain is transferred without distortion to amplify the thermal wave. The amplitude of the thermal wave can be modeled by a linear strain-dependence reaching a maximum variation of ~ ± 0.2°C (10 times over the accuracy) (Fig 1e). The linear relationship suggests that the thermal mechanism is occurring as soon as the smallest measurable shear strain values; i.e. inherent to the liquid deformation. Fig.1e insert displays the rate of the temperature variation over the strain. $\Delta T/\gamma$ is equivalent to the "gain" and measures the efficiency (quality) of the energy conversion. This "gain" is nearly constant for the three bands (insert of Fig.1e). Another remarkable feature is the symmetry of the evolution of the cold and hot bands pointing out a rigorous compensation to keep the net temperature unchanged.

At 1 rad/s, the strain dependence of the thermal signal is more complex. Two thermal regimes are observed depending on the strain amplitude:

- At relatively "low" strain amplitudes ($\gamma$ < 2500% at $\omega$ = 1 rad/s), the thermal modulations obeys a sin wave with a phase shift smaller than π/4 in advance with the applied strain wave (Fig.2a). Specifically, the phase shift at $\gamma$ = 1000% is $\delta\varphi_{middle}$ ~ 32° ± 2°, $\delta\varphi_{bottom}$ ~ 29° ± 2° for the cold and the hot bands respectively. The liquid exhibits an output thermal response synchronous with the input strain wave. The "hot" and "cold" waves evolves symmetrically confirming that the system works nearly in an exact thermal compensation as the strain amplitude increases.

- At higher strain amplitude of the strain wave ($\gamma$ > 2500% at $\omega$ = 1 rad/s), the linear relationship with the input strain wave does not hold anymore, meaning that non-linear effects start to take place. Fig.1f shows the progression of the amplitude of both hot and cold wave amplitudes. For the highest strain values, the thermal variation no more evolve as shear strain increases sketching the onset of a plateau (in agreement with a decrease of the gain $\Delta T/\gamma$ at high strain amplitudes illustrated in the insert of Fig.1f). Simultaneously the shape of the thermal wave evolves and deforms. Fig.2 details the evolution of the cold wave. The wave splits in harmonics, while increasing the thermal amplitude for higher strain values. The extrapolation to infinite strain amplitudes is a contraction of the thermal amplitude and possibly a vanishing of the thermal variation (experimental restrictions do not allow us to go further). The infinite strain corresponds to steady-state conditions; i.e. to the flow regime. The vanishing of the thermal behavior at infinite strain is in agreement with the common observation of absence of noticeable thermal effects at macroscopic scale.

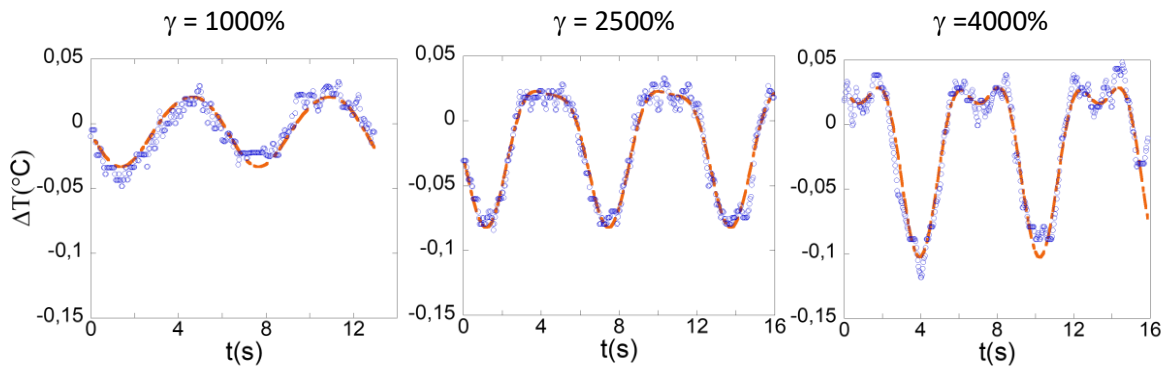

**Figure 2.** Evolution of the cold thermal wave (middle band) versus shear strain at constant frequency (1 rad/s) for 1000%, 2500% and 4000% shear strain amplitude. At large strain, the thermal wave exhibits harmonics with a smoothed amplitude. Orange dotted lines are fits following eq. $f(t) = A_0 + \sum_{i=1}^{2} A_i \sin(\omega_i t + \varphi_i)$.

## 1.3 Frequency dependence of the thermal signal:

Fig.3a-3c shows the influence of the frequency on the thermal wave in a range of 0.5 - 5 rad/s (measured at the maximum strain wave amplitude $\gamma = 4000\%$). At low frequency, the thermal waves reproduce the input strain waveform (Fig.3a illustrates the cold band at $\omega = 0.5$rad/s). As the frequency is increased, the thermal wave progressively loses the input sinus shape (Fig.3b and 3c) and its amplitude decreases of nearly 40% from 0.5 to 5 rad/s. The thermal signal is decomposed using Fast Fourier Transformation (FFT) [16] to compare its frequency domain with the one of the applied shear strain signal. FFT analysis (Fig.3d-3f) shows that higher order harmonics appear in addition to the fundamental frequency by increasing the frequency (Fig.3b 3d-f). The thermal waves are fitted using the sum of the harmonics: $f(t) = A_0 + \sum_{i=1}^{n} A_i \sin(\omega_i t + \varphi_i)$, where $i$ is a numerator and $n$ represents the number of harmonics used, $A_i$ constant and $\varphi_i$ the phase shift. The generation of thermal harmonics implies that the excitation frequency is faster than the relaxation mechanism between two successive oscillations and so, the thermal response is no more stable.

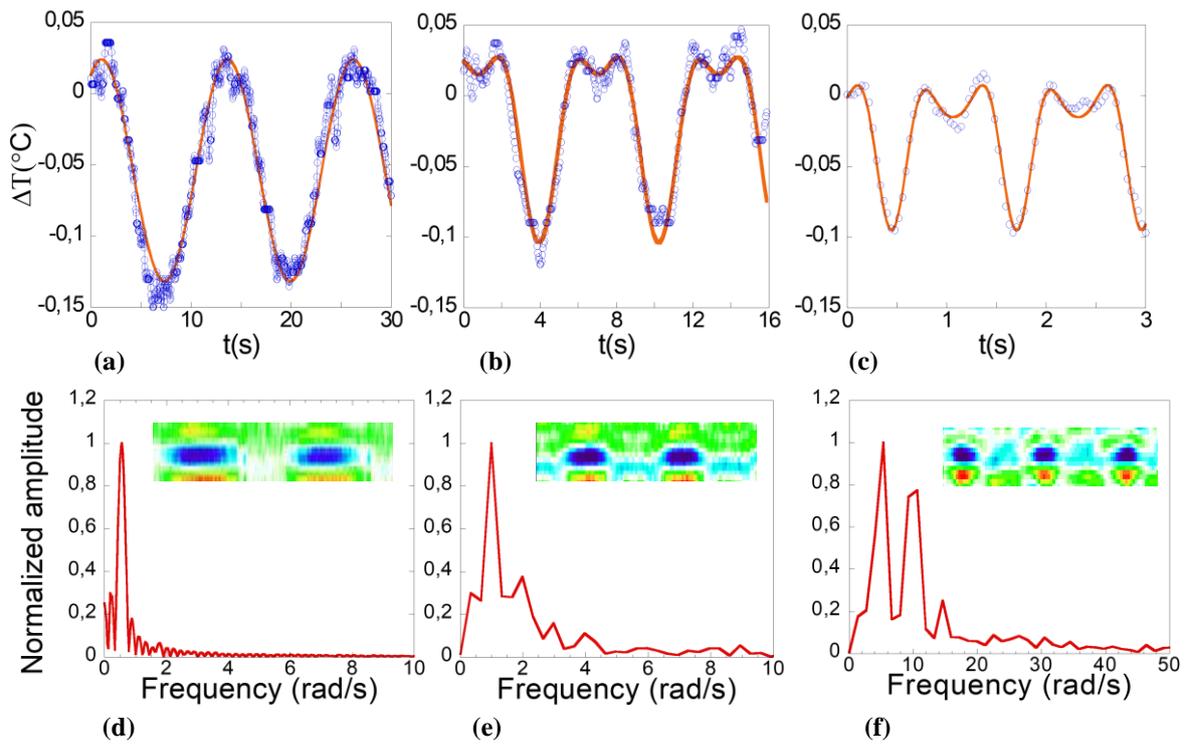

**Figure 3.** Influence of the frequency on the evolution of the cold thermal wave at large strain amplitude ($\gamma = 4000\%$) and corresponding FFT analysis. Increasing the frequency produces thermal harmonics and a collapse of the thermal variation. Data points (blue circles) corresponds to the thermal variation ΔT within PPG-4000 middle band at 0.240mm, , **a)** $\omega = 0.5$ rad/s, **b)** $\omega = 1$ rad/s, **c)** $\omega = 5$ rad/s. Orange lines are fitting of eq. $f(t) = A_0 + \sum_{i=1}^{n} A_i \sin(\omega_i t + \varphi_i)$ for a) n=1 b) n=2 c) n=3, Fourier Transform signal of the waves depicted of **d)** Fig. 3a, **e)** Fig. 3b, **f)** Fig. 3c. Inset of each graph corresponds to the equivalent 2D thermal mapping.

## 2. Probing van der Waals liquid:

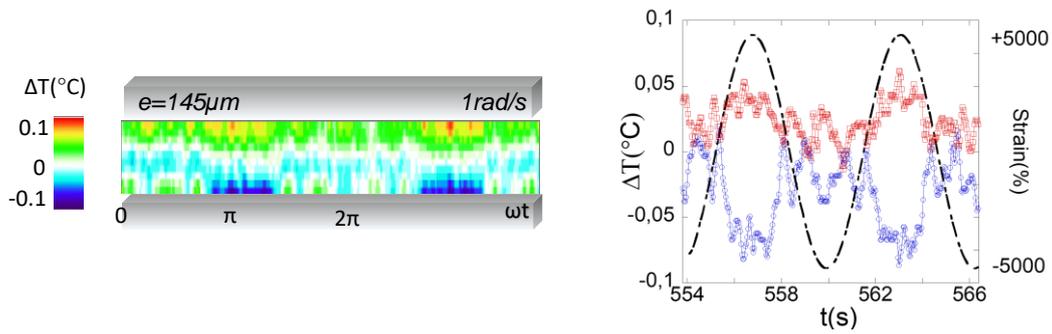

**Figure 4.** Thermal response of pentadecane (0.145mm, 1rad/s, 5000%). Red and blue points represent top and bottom band respectively. Black dotted line represents the applied strain – alumina substrate, room temperature measurements.

Figure 4 demonstrates that the thermal response is also measurable for a very low molecular weight liquid; the pentadecane. We observe a weak thermal response for low gap thickness and high shear strain values with the same characteristics as the PPG, namely the generation of hot and cold bands. Specifically, thermal variation is not observed in the whole liquid, but near the walls (30 – 40 μm width). The middle part of pentadecane gap, which is less effected than the surfaces, does not exhibit thermal oscillations. The weaker thermal response could be attributed to the non-polar nature of pentadecane, which is a van der Waals molecule while PPG is a hydrogen bonded oligomer. Previous studies report on liquid shear elasticity is higher and more easily identified in polymer melts [4] than less viscous liquids (glycerol, water) [5-6], implying that different cohesion of the liquids matter on the intensity of solid-state properties in liquids. Another possible cause of different magnitude of the thermal effect could be that the liquids have different natural frequencies. On the other hand, the identification of thermal response in a non-polymeric liquid suggests that the observed thermo-elastic effect is generic to liquids.

## 3. Discussions-conclusions:

The present results show that the thermal approach of solids is perfectly adaptable for the study of mesoscopic liquids. We have established that liquids like the PPG emit a synchronous hot and cold thermal modulation upon applying an oscillatory shear strain wave. The examination of the thermal wave versus frequencies and strain amplitude showed that the thermal response is a stable sin wave within a low frequency range ($\leq 0.5$ rad/s) at small gap (250µm). A nearly linear variation of the temperature variation is observable for both cold and hot thermal bands at moderate shear strain amplitude (Fig.1e-1f). The thermal waves exhibit a slight phase shift (less than $\pi/4$) with respect to the input strain wave, indicating that a part of the mechanical energy is dissipated. At frequencies over 1 rad/s or at very large shear strain amplitudes, a distortion of the thermal wave indicates that the response is no more stable and exhibits harmonics. The generation of thermal harmonics suggests that further dissipation takes place due secondary effects and an applied excitation faster than the relaxation time of the thermal effect (thus estimated at $\tau \sim 1-2s$). We identify odd and even harmonics for the thermal response as the strain and/or frequency increases. The extrapolation at high frequency or at high strain amplitude indicates a smoothing of the thermal oscillation towards a flat signal (constant temperature).

The thermal behavior reveals that apparent "viscous" liquid behaviors hide a more complex mesoscale mechanism. The thermal analysis reveals the liquid ability to convert the shear wave energy in local thermodynamic states. Such effect has characteristics of a conservative mechanism that are: - the linearity as the shear strain increases, - the synchronism and a nearly instant thermal responses to the strain wave, and at the probed frequencies, an absence of heat conduction between bands and with the environment. These properties might be regarded as mechanically induced adiabatic thermodynamic events (in the linear domain). These are typical of a thermoelastic effect; i.e. mediated by a liquid elasticity varying as a function of temperature. Shear elasticity was indeed identified in mesoscale liquids by stress measurements and depending on the scale considered. Since the early works of Derjaguin in 1989 [3], evidence of the existence of long range interactions in liquids accumulates from both experimental results [3-7, 17-19] and fundamental models. The later ones come from new theoretical developments [20, 9-13], inspired from solids and revisiting the Frenkel theory of propagation of transverse waves in liquids [21]. A scale dependence of the low frequency shear modulus scales is also predicted as $G' \sim h^{-3}$ [12, 21]. This might explain why elastic effects become dominant as the scale decreases [3-7] while high resolved inelastic measurements are now able to detect vibrational patterns such as shear phonon excitations in self-assembly liquids [23]. The present thermal results are an indirect proof of the shear wave propagation and shows that the accumulated shear energy modifies the liquid thermodynamic at small scale (<500μm). From a mechanical point of view, reversible temperature changes induced by the deformation known as thermo-(shear)elastic effect are forbidden by the conventional liquid incompressibility assumption. Non-zero thermoelastic coefficients (here estimated at $K_{shear} = \Delta T/\gamma$ about $3.6 \pm 0.3 \cdot 10^{-3}$ K$^{-1}$) imply that isobaric expansion and isothermal compressibility of liquids are also non-zero, opening great avenues for future developments in the emerging fields of nano and microfluidics.

**Corresponding Author:** Correspondence should be addressed to laurence.noirez@cea.fr

**Author Contributions:** The manuscript was written through contributions of all authors. E.K. undertook the experimental work and wrote the paper. L.N. proposed the scientific subject, supervised the developments and contributed to the writing.

**Acknowledgments:** The authors would like to thank P. Baroni for instrumental innovation and assistance.

**Funding Sources:** This work has received funding from the European Union's Horizon 2020 research and innovation programme under the Marie Sklodowska-Curie grant agreement No 766007 and from the LabeX Palm (ANR-11- Idex-0003-02).